\documentclass[a4paper]{article}
\usepackage{ISCSLP2024}
\usepackage{ifthen}
\usepackage{subfigure}
\usepackage{caption}

\newboolean{blind}
\title{Exploring the Role of Audio in Multimodal Misinformation Detection}
\name{
	\ifthenelse{\boolean{blind}}{Anonymous to ISCSLP}
	{Moyang Liu$^1$, Yukun Liu$^2$$^{\ast}$\thanks{*Corresponding author}, Ruibo Fu$^3$, Zhengqi Wen$^4$, Jianhua Tao$^4$, Xuefei Liu$^3$, Guanjun Li$^3$}
}
\address{
  \ifthenelse{\boolean{blind}}{Anonymous to ISCSLP}
  {
  $^1$Beihang University\\
  $^2$School of Artificial Intelligence, University of Chinese Academy of Sciences\\
  $^3$Institute of Automation, Chinese Academy of Sciences\\
  $^4$Beijing National Research Center for Information Science and Technology, Tsinghua University
  }
}

\email{
	\ifthenelse{\boolean{blind}}{Anonymous to ISCSLP}
	{moyang\_liu@buaa.edu.cn, yukunliu927@gmail.com}
}

\begin{document}

\maketitle
\begin{abstract}
With the rapid development of deepfake technology, especially the deep audio fake technology,
misinformation detection on the social media scene meets a great challenge.
Social media data often contains multimodal information which includes audio, video, text, and images. However, existing multimodal misinformation detection methods tend to focus only on some of these modalities, failing to comprehensively address information from all modalities.
To comprehensively address the various modal information that may appear on social media, this paper constructs a comprehensive multimodal misinformation detection framework. 
By employing corresponding neural network encoders for each modality, the framework can fuse different modality information and support the multimodal misinformation detection task.
Based on the constructed framework, this paper explores the importance of the audio modality in multimodal misinformation detection tasks on social media. 
By adjusting the architecture of the acoustic encoder, the effectiveness of different acoustic feature encoders in the multimodal misinformation detection tasks is investigated. 
Furthermore, this paper discovers that audio and video information must be carefully aligned, otherwise the misalignment across different audio and video modalities can severely impair the model performance.
\end{abstract}

\section{Introduction}

The rapid development of social media has provided a broad platform for the dissemination of misinformation, which has posed substantial challenges to society\cite{zhou2020survey}. Unlike traditional news media, social media allows anyone to become an information publisher, greatly increasing the diversity of information sources\cite{zhang2020overview}. However, this leads to the proliferation of misinformation. Moreover, with the continuous development of artificial intelligence generative content(AIGC)\cite{zhou2019network}, people can easily create realistic images, audio, and video content, reducing the cost of producing forgeries and making them harder to detect\cite{singhal2019spotfake}. Given that misinformation can cause panic, hatred, social unrest, and even affect political elections and public policy making, effective measures are needed to address this issue\cite{shu2019beyond}. 

Therefore, to assist people in identifying and distinguishing between authentic and false content, the detection of misinformation becomes particularly important\cite{shu2017fake}. In recent years, there has been a significant amount of research focusing on misinformation detection\cite{shu2017exploiting}. Initially, research on misinformation detection predominantly focused on unimodal approaches, such as detecting scam text messages or identifying fake voice recordings\cite{perez2017automatic}. Reis et al.\cite{reis2019supervised} extracted various textual features from news articles, including grammatical characteristics, lexical functions, psycholinguistic features, semantic features, and subjectivity, to identify their veracity\cite{yang2018ti}. In the field of speech, Subramani et al.\cite{subramani2020learning} present four parameter-efficient convolutional architectures for fake speech detection.   

However, unimodal misinformation detection suffers from limited available content and fails to meet real-life needs\cite{oshikawa2018survey}. Since contemporary fake news typically contains information across multiple modalities\cite{zhou2020fake}, fully utilizing information from each modality often leads to a better understanding of the news content and thus more effective detection\cite{parikh2018media}.  Kai et al.\cite{nakamura2019r} constructed a large-scale multimodal fake news dataset called Fakeddit, which includes text, images, metadata, and comments from the social media platform Reddit, comprising over one million samples. Jing et al.\cite{jing2023multimodal} proposed a progressive fusion network MPFN for multimodal disinformation detection, which captures the representational information of each modality at different levels. Nevertheless, most existing multimodal models have neglected certain modalities, which hinders their performance improvement in misinformation detection\cite{zhou2019fake}. Qi et al.\cite{qi2023fakesv} constructed China's largest fake news short video dataset FakeSV, which includes various contents such as titles, videos, keyframes, audios, metadata, and user comments. 

There are currently many issues in multimodal misinformation detection. Firstly, in the research of multimodal models, the audio modality is often undervalued or ignored\cite{zhou2021joint}. This oversight results in models that are unable to process information comprehensively and accurately, thus reducing the accuracy and reliability of misinformation detection\cite{wang2017liar}.
Secondly,  further exploration is needed to determine which type of acoustic encoder is most suitable for the multimodal misinformation detection task, which makes significant sense for future research\cite{yang2019unsupervised}.
Finally, there is the problem of modality alignment\cite{shu2022see}. For dynamic modalities such as audio and video, it is essential to consider temporal alignment; otherwise, it can adversely affect the model's performance\cite{hong2019learnable}.

To explore the role of audio in multimodal misinformation detection,
we construct a comprehensive multimodal misinformation detection framework that can simultaneously handle the audio, visual, and textual information.
Extensive experiments on the framework demonstrate the importance of the audio modality.
By comparing different acoustic encoders under the framework, we explore the applicability of different acoustic encoders in the multimodal misinformation detection task.
With the modality fusing between audio, video, and social information, we discovered the detrimental impact of their misalignment on the multimodal misinformation detection task, which provides valuable insights for future research.

\begin{figure*}[h]
    \centering
    \includegraphics[width=15cm]{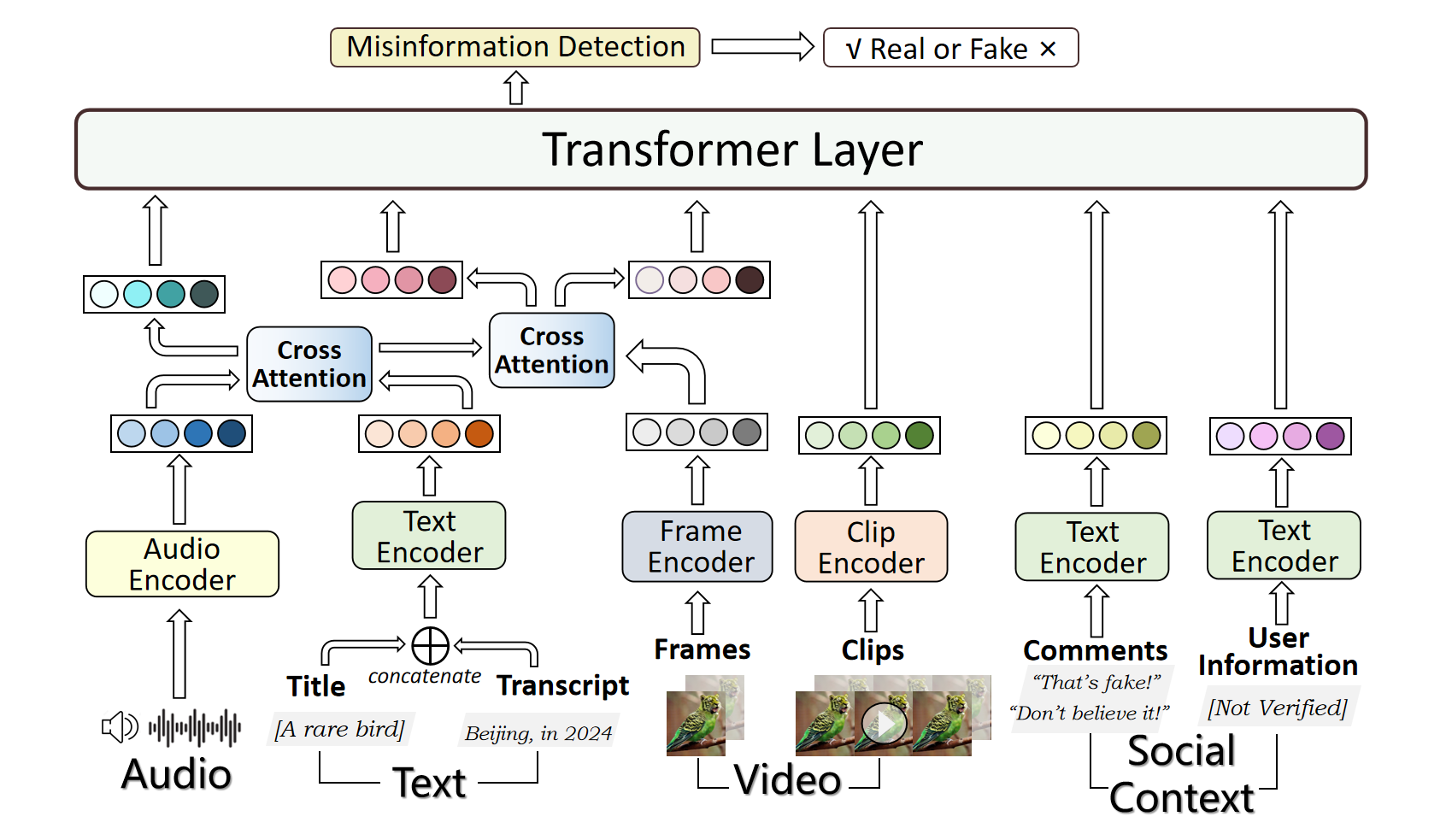}
    \caption{Architecture of the constructed multimodal misinformation detection framework }
    \label{f1}
\end{figure*}

\section{Methods}

\subsection{Framework}

To tackle the multimodal misinformation detection task and cover modalities that appear on social media, we construct a multimodal misinformation detection framework based on the work of Qi et al\cite{qi2023fakesv} as shown in Figure \ref{f1}. 

In the framework, to address the proliferation of various types of misinformation encountered in practice, we utilize multiple modalities of short video content (text, audio, image, video, and social context) to comprehensively extract useful information. 
For each modality, we employ a modality-specific encoder to extract features and integrate the extracted modal information for misinformation detection.
Cross-attention\cite{vaswani2017attention} is a mechanism that allows a model to focus on relevant parts of different input sequences simultaneously. It enables the model to learn the relationships and dependencies between different modalities, thus enhancing the quality of the combined representation.

Subsequently, to understand the content of the information from different levels, we used a cross-attention layer to fuse the features extracted from the text and audio modalities. Next, we fused the features obtained from the cross-attention layer with the features extracted from the video modality. This process enables a more comprehensive understanding of the information and allows for complementary information integration. Finally, the fused features from each modality are fed into a transformer\cite{vaswani2017attention} layer, and the output of this layer is classified into true or false categories by a classifier.

\begin{table*}[t]
\centering
\setlength{\belowcaptionskip}{0.2cm}
\begin{tabular}{ccccccccc}
\hline
Audio Encoder &Audio  &  Text      & Video     & Social &Accuracy & F1-score &Precision & Recall \\ \hline         
w/o Audio        &                   & \checkmark         &           &            & 75.80                     & 75.68                     & 75.63                      & 75.88                   \\
VGG         &\checkmark                   & \checkmark         &           &            & 81.55                     & \textbf{81.40}            & 81.34                    & \textbf{81.73}          \\
wav2vec2.0     &\checkmark                & \checkmark         &           &            & \textbf{82.29}            & 81.03                     & \textbf{83.44}             & 81.21                   \\ \hline
w/o Audio        &                   & \checkmark         & \checkmark         &            & 77.02                     & 76.89                     & 76.83                      & 77.15                   \\
VGG          &\checkmark                  & \checkmark         & \checkmark         &            & \textbf{80.92}            & \textbf{80.42}            & \textbf{80.34}             & \textbf{81.24}          \\
wav2vec2.0    &\checkmark                 & \checkmark         & \checkmark         &            & 78.60                     & 78.46                     & 78.40                      & 78.78                   \\ \hline
w/o Audio          &                 & \checkmark         & \checkmark         & \checkmark          & 79.84                     & 79.72                     & 79.68                      & 79.80                   \\
VGG           &\checkmark                 & \checkmark         & \checkmark         & \checkmark          & 80.44                     & 80.18                     & 80.13                      & 80.24                   \\
wav2vec2.0     &\checkmark                & \checkmark         & \checkmark         & \checkmark          & \textbf{81.18}            & \textbf{80.93}            & \textbf{80.88}             & \textbf{80.99}          \\ \hline
\end{tabular}
\caption{The experimental results of the multimodal disinformation detection framework using different configurations of audio encoders and combinations of modalities (text, video, social context).}
\label{tabel 1}
\vspace{-2.0em}
\end{table*}

\subsection{Feature Encoder}

To extract valuable information from each modality, it is crucial to select an appropriate feature encoder based on the current modality.

\textbf{Text:}

Textual information in short videos on social media often includes titles and transcripts. Since titles are typically limited in length, we extracted the video transcripts and concatenated them with the titles to enhance completeness. Then, these concatenated textual information are fed into a pre-trained BERT model\cite{devlin2018bert} to extract textual features.

\textbf{Video:}

To detect the spatiotemporal and multi-granularity information in the video modality, we captured different features from both frame and clip levels.

a. Frame At the frame level, we extracted a specific number of frames from each video and fed them into a pre-trained VGG19 model to learn static visual features.

b. Clip At the clip level, we extracted 16 consecutive frames centered around each time step as a video segment and used a pre-trained C3D model\cite{tran2015learning} to extract motion features. We then applied an averaging operation to obtain aggregated motion features.

\textbf{Social context:}

Social context includes user information and comments. Similar to the processing method for the textual modality, we fed this information into a pre-trained BERT model to extract features separately.

\textbf{Audio: } 

With the advancement of deep learning techniques in the audio domain, several mature acoustic encoders have been developed. To investigate the application of different acoustic encoders in multimodal misinformation detection, we employed the CNN-based VGG\cite{simonyan2014very} encoder and the Transformer-based wav2vec2.0\cite{baevski2020wav2vec} encoder for extracting acoustic features.

\begin{figure}[htbp]
\centering
\subfigure[VGG Encoder]{\label{fig:subfig1}\includegraphics[width=0.45\textwidth]{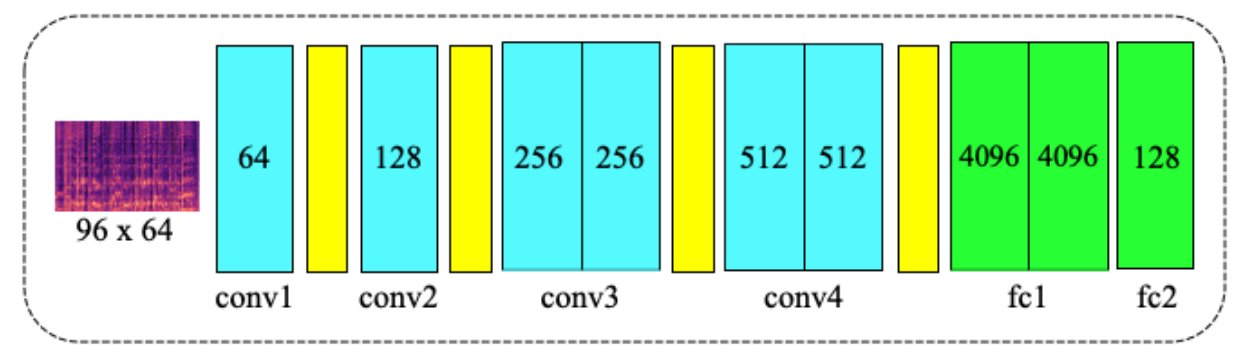}}
\subfigure[Wav2vec2.0 Encoder]{\label{fig:subfig2}\includegraphics[width=0.45\textwidth]{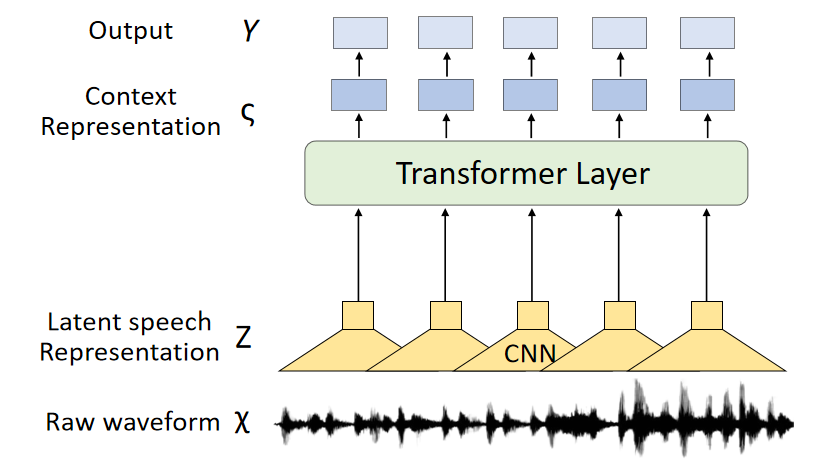}}
\caption{Acoustic Feature Encoders: (a) is the architecture of VGG, which is composed of a series of CNN blocks. (b) is the architecture of wav2vec2.0, it can obtain more accurate acoustic representations in the unsupervised way.}
\label{fig:mainfig}
\end{figure}

The VGG encoder is a convolutional neural network used for audio feature extraction. As shown in Figure \ref{fig:subfig1}, its basic building blocks are convolutional layers and pooling layers. These convolutional layers are typically employed to process the time-frequency representations of audio (such as Mel spectrograms\cite{shen2018natural} or spectrograms), extracting high-level audio features.

And wav2vec2.0 is a self-supervised speech representation learning model, consisting of three main components: a Convolutional Neural Network (CNN) encoder, a quantization module, and a Transformer encoder. Figure \ref{fig:subfig2} displays the architecture of wav2vec2.0. It leverages self-supervised learning to learn speech representations from large amounts of unlabeled audio data and has already demonstrated excellent performance in various speech tasks

\subsection{Feature Fusion}

The fusion of features across modalities enhances the completeness of information and deepens the model's understanding of semantics. To achieve this, we used three cross-modal transformer layers to strengthen the interaction and integration between different modalities.

Audio features are first fused with textual features to complement each other. Denote that the output from the text encoder as \( F_t = (x_1, x_2, \ldots, x_n) \), where \(x_i  \) represents the feature of the i-th word and l is the text length. And the extracted audio features are denoted as \( F_a = (a_1, a_2, \ldots, a_n) \), where n is the number of audio frames. From \( F_t \) generate the query matrix \( Q_t \), and from \( F_a\) , generate the key matrix \( K_a \) and the value matrix \( V_a \) . Subsequently, calculate the attention weights and compute the attention-weighted sum of the text features based on the audio features.

\(F_{t \leftarrow a} = \text{Attention}(Q_t, K_a, V_a) = \text{softmax}\left(\frac{Q_a K_t^T}{\sqrt{d_k}}\right) V_t\)

where \( d_k \) is the dimensionality of the key vectors. Similarly, from \( F_a \), generate the query matrix \( Q_a \), and from \( F_t \), generate the key matrix \( K_t \) and the value matrix \( V_t \). Calculate the attention-weighted sum of the audio features based on the text features as follows:

\(F_{a \leftarrow t} = \text{Attention}(Q_a, K_t, V_t) = \text{softmax}\left(\frac{Q_t K_a^T}{\sqrt{d_k}}\right) V_a\)

Then, we feed the audio-enhanced textual features \(F_{t \leftarrow a}\) and frame features \(F_f\) into the second transformer layer, resulting in textual features enhanced by audio and visual frames \(F_{t \leftarrow a,f}\), as well as text-enhanced frame features 
\(F_{f \leftarrow t}\). Then we average them and obtain the features \( x_t \) , \( x_a \), and \( x_f \).

Finally, we use self attention mechanism to enhance the relevance between content features and social context features. We concatenate the obtained six features into a sequence and input them into a transformer layer to obtain the final multimodal fused features.

\section{Experiments}

\subsection{Dataset}

We utilized the FakeSV dataset constructed by Qi et al.\cite{qi2023fakesv}, which comprises a large collection of Chinese news short videos. This dataset includes multiple modalities such as text, video, audio, and social context, which can cover the data from various modalities in social media scenarios.

\subsection{Experiment Details}

Consequently, the dataset was divided into training, validation, and test sets in a 70:15:15 ratio following a chronological order. The model utilized the cross-entropy loss function and AdamW\cite{loshchilov2017decoupled} optimizer, with a batch size of 64. 
The final results were obtained by evaluating this best model on the test set.
BERT is employed as the text encoder.
VGG19 and C3D are employed as the video encoder, and the social encoder also adopts BERT.
VGG and wav2vec2.0 are compared to which is the better acoustic encoder.

\subsection{Performance comparison}

We investigated the contribution of the audio modality to the overall model performance. In typical multimodal studies, audio features are often overlooked or underutilized. However, our experiments demonstrate that audio features play a crucial role in detecting misinformation. Specifically, the semantic information carried in the audio can provide additional context and cues that are not present in other modalities.

To quantify the impact of the audio modality, we conducted experiments comparing model performance with and without the inclusion of audio features. As shown in Table \ref{tabel 1} and Table \ref{tabel 2}, the results clearly indicate that the model's ability to detect fake news improves significantly when audio features are included. This improvement underscores the importance of considering audio information, which captures nuances in speech, tone, and background sounds that are critical for understanding and identifying misleading content.

\begin{table}[h]
\centering
\setlength{\belowcaptionskip}{0.2cm}
\begin{tabular}{ccccc}
\hline
Audio & Text & Video & Social & Accuracy       \\ \hline
      & \checkmark    & \checkmark     & \checkmark      & 79.84          \\
\checkmark     & \checkmark    & \checkmark     & \checkmark      & \textbf{81.18} \\
\checkmark     & \checkmark    &       &        & \textbf{82.29} \\
\checkmark     & \checkmark    & \checkmark     &        & 78.60          \\
\checkmark     & \checkmark    &       & \checkmark     & \textbf{80.87} \\
\checkmark     &      & \checkmark     & \checkmark      & 78.60          \\ \hline
\end{tabular}
\caption{The experimental results illustrate the contribution of the audio modality to the model. Audios can provide additional context and cues that are not present in other modalities}
\label{tabel 2}
\vspace{-2.0em}
\end{table}

\subsection{Further Exploration}

We compared the performance of two acoustic encoders, VGG and wav2vec 2.0, to evaluate their effectiveness in extracting audio features for multimodal fake news detection. As shown in Table \ref{tabel 3}, our experiments revealed that the features extracted by the wav2vec model integrated more effectively with features from other modalities, resulting in better overall model performance.

\begin{table}[h]
\centering
\setlength{\belowcaptionskip}{0.2cm}
\begin{tabular}{ccccc}
\hline
Audio Encoder & Text & Video & Social & Accuracy       \\ \hline
VGG           & \checkmark    &       &        & 81.55          \\
wav2vec2.0    & \checkmark    &       &        & \textbf{82.29} \\ \hline
VGG           & \checkmark    & \checkmark     &        & \textbf{80.92} \\
wav2vec2.0    & \checkmark    & \checkmark     &        & 78.60          \\ \hline
VGG           & \checkmark    &       & \checkmark      & 80.63          \\
wav2vec2.0    & \checkmark    &       & \checkmark      & \textbf{80.87} \\ \hline
VGG           &      & \checkmark     & \checkmark      & 77.84          \\
wav2vec2.0    &      & \checkmark     & \checkmark      & \textbf{78.60} \\ \hline
VGG           & \checkmark    & \checkmark     & \checkmark      & 80.44          \\
wav2vec2.0    & \checkmark    & \checkmark     & \checkmark      & \textbf{81.18} \\ \hline
\end{tabular}
\caption{Comparison of two audio encoders. The features extracted by the wav2vec model integrated more effectively with features from other modalities, resulting in better overall model performance.}
\label{tabel 3}
\vspace{-2.0em}
\end{table}

This improvement can be attributed to the nature of the features each model extracts. The VGG model is more adept at capturing background sounds, which, while useful in some contexts, may not provide the semantic depth needed for accurate fake news detection. On the other hand, wav2vec excels at extracting semantic-level features from audio, which align more closely with the textual features processed by BERT models. This semantic alignment facilitates more effective feature fusion and enhances the model's ability to understand and analyze the content. Moreover, detecting fake news often requires a deep understanding of the spoken content within videos, as nuances in speech and verbal cues can be critical indicators of authenticity. The wav2vec model's ability to capture these semantic details makes it particularly suitable for this task. Consequently, our results indicate that incorporating wav2vec as the audio encoder significantly improves the model's performance in detecting misinformation.

Additionally, the experimental results highlight the impact of insufficient alignment between modalities on model performance. As shown in Table \ref{tabel 4}, the inclusion of the video modality and social modality did not enhance the model's performance; in fact, it led to a decline. This performance drop is attributed to the inadequate alignment between the different modalities.

\begin{table}[h]
\centering
\setlength{\belowcaptionskip}{0.2cm}
\begin{tabular}{ccccc}
\hline
Audio & Text & Video & Social & Accuracy       \\ \hline
\checkmark     & \checkmark    &       &        & \textbf{82.29} \\
\checkmark     & \checkmark    & \checkmark     &        & 78.60          \\
\checkmark     & \checkmark    &       & \checkmark      & 80.87          \\
\checkmark     & \checkmark    & \checkmark     & \checkmark      & 81.18          \\ \hline
\end{tabular}
\caption{The experimental results show the impact of insufficient alignment between modalities on model performance}
\label{tabel 4}
\vspace{-2.0em}
\end{table}

Effective multimodal integration requires that the features extracted from different modalities are properly aligned and complementary to each other. However, if the features are misaligned or if the integration mechanism is not robust, the additional information from one modality can introduce noise rather than valuable insights, thus deteriorating the overall model performance. In our case, the video features, when not adequately aligned with the audio and text features, contributed to a decline in accuracy.

\section{Conclusions}

In this study, we addressed the challenge of detecting misinformation in short videos using a multimodal approach, leveraging the rich and diverse features provided by different modalities. Our experiments on the FakeSV dataset demonstrated the significant potential and challenges of multimodal misinformation detection. 

Our research revealed that audio features play a critical role in improving the model's accuracy, as they provide valuable context that aids in distinguishing between true and false content. Additionally, we found that the wav2vec model, which excels in capturing semantic-level audio features, significantly enhances fake news detection when compared to the VGG model. However, we also observed that improper alignment between different modalities, particularly with video features, can negatively impact performance. This highlights the importance of robust alignment mechanisms to ensure that multimodal features complement each other effectively.

Overall, our study demonstrates that a well-designed multimodal framework can significantly enhance the detection of misinformation in short videos. Future work should focus on improving the alignment and fusion of multimodal features to further boost performance. Additionally, expanding the dataset to include more diverse and representative samples can help in building more generalized and robust models.

\section{Acknowledgements}

This work is supported by the National Natural Science Foundation of China (NSFC) (No.62101553, No.62306316, No.U21B20210, No.62201571).

\bibliographystyle{IEEEtran}

\bibliography{article}


\end{document}